# Loss Aversion and State-Dependent Linear Utility Functions for Monetary Returns

**Somdeb Lahiri**

**(Former Professor) PD Energy University, Gandhinagar (EU-G), India.**

**ORCID: https://orcid.org/0000-0002-5247-3497**

**somdeb.lahiri@gmail.com**

**Abstract**

We present a theory of expected utility with state-dependent linear utility functions for monetary returns, that incorporates the possibility of loss-aversion. Our results relate to "first order stochastic dominance", "mean-preserving spread", "increasing-concave linear utility profiles" and "risk aversion". As an application of the expected utility theory developed here, we analyze the contract that a monopolist would offer in an insurance market that allowed for partial coverage of loss.

**Keywords:** money, utility, state-dependent linear, loss aversion, first order stochastic dominance, mean-preserving spread, risk aversion

**JEL Classifications:** D81.

## 1. Introduction:

A common argument against linear utility function for monetary returns is, that an agent with such a utility function would have no incentive to insure himself against possible "loss". However, this argument seems to collapse if the linear utility function for monetary returns is state dependent and the "probability of the gain or loss" is spelt out as the "probability of the **state of nature** (**SON**) in which there is the gain or loss" with the constant marginal utility of monetary returns in the "worse" state being more than the constant marginal utility of money in the "better" state. In what follows we will refer to **states of nature (SONs)**.

The seminal contribution of Kahneman and Tversky (Kahneman and Tversky 1979) noted the experimentally verified observation that agents tend to have a marginal utility of loss that is no less- if not higher-than the marginal utility of gain, so that a typical utility function for monetary returns $u:\mathbb{R}\to \mathbb{R}$ may be of the form $u(x) = u^+\max\{x,0\} + u^-\min\{x,0\}$ with $u^- \geq u^+ > 0$. This phenomenon is known as "loss aversion". Thus, any utility function of this form can be represented by a pair of real numbers $(u^-, u^+)$ where $u^- \geq u^+ > 0$. Allowance is made for the possibility of $u^- = u^+$. It is generally assumed that under normal circumstances $u^- - u^+$ is a non-decreasing function of initial wealth, thereby implying that wealthier individuals are more "risk averse" than those individuals who are less wealthy.

The dominant interpretation of probability in expected utility theory, is the one due to Ramsey and de Finetti. Brief discussions along with intuitive motivation of such probabilities are available in two recent papers (Lahiri 2023a, Lahiri 2023b). The Ramsey-de Finetti subjective probability of an "event" or "state of nature" (say E) that is assessed by an agent is the price (say p) that the agent would we willing to pay for a simple bet that returns one unit of money if state of nature 'E' occurs and nothing otherwise, so that the expected monetary value of the simple bet to the agent is zero. Thus, if the average utility of money in state of nature E is a constant, say $\mu > 0$, then for one unit of money in state of nature E, the agent will be willing to forego $\mu p$ units of utility and for $\xi$ units of money in state of nature E the agent will be willing to forego $\mu p \xi$ units of utility, the latter being the utility the agent willingly forgoes for $\xi$ simple bets of the type we have just discussed. $\xi$ simple bets, each of which returns one unit of money if E occurs and nothing otherwise, is identical to a bet that returns $\xi$ unit of money if E occurs and nothing otherwise. Thus, Ramsey-de Finetti subjective probability fits comfortably with "expected utility theory" based on constant average state dependent utility. On the other hand, if the average utility in state of nature E is "non-constant", then there exists $\xi$ such that the average utility of $\xi$ units of money is not equal to the average utility of $p\xi$ units of money. For a bet that returns $\xi$ units of money in state of nature E and nothing otherwise, the agent will be "willingly foregoing" the utility of $p\xi$ units of money and not 'p' times the utility of $\xi$ units of money, the latter being the expected utility of the bet to the agent. Hence, on the face of it, there seems to be a mismatch between Ramsey-de Finetti subjective probability and expected utility theory based on such an interpretation, if state-dependent average utility of money is "non-constant".

In this paper we attempt a reconciliation of Ramsey-de Finetti subjective probability with the kind of loss aversion that Kahneman and Tversky had suggested, by allowing an agent to sell a "simple bet on an event" at a higher price than what the agent would be willing to pay to buy it. Perhaps this price wedge reflects a "transaction cost" that the seller incurs. After all, transaction costs are not incurred in "thought experiments"- including the ones used for the purpose of evaluating Ramsey-de Finetti subjective probabilities. Alternatively, it could be the "normal profit" or "the opportunity cost of the seller's labour time" as in microeconomics. Whether justified or not, such an argument is one way in which the conflict between "loss aversion" and Ramsey-de Finetti subjective probabilities can be resolved. This requires invoking "state-dependent linear utility functions for monetary gains and losses" while allowing for the average utility of losses to exceed the average utility of gains. This allows, decision analysis based on expected utility to fit meaningfully with the concept of expected utility based on Ramsey-de Finetti probabilities as well as loss aversion, keeping in mind the "caveat" for the price wedge in terms of transaction costs or normal profits that we discussed earlier. A comprehensive exposition of the early stages of the analysis of decision making under uncertainty with state dependent preferences is available in the work of Karni (Karni 1985).

In the next section of the paper, we provide a motivation for our discussion in the subsequent sections, by considering a "toy model" of insurance against a risky loss. We apply expected state dependent linear utility analysis in this model and show that insurance is possible under state-dependent "risk neutrality". In the third section, we present the formal framework for "expected utility theory with state-dependent linear utility functions for monetary returns". Using concepts introduced in this section, in subsequent sections we introduce "first order stochastic dominance", "mean-preserving spread", "increasing-concave linear utility profiles" and "risk aversion". As an application of the expected utility theory developed here, we analyze the contract that a monopolist would offer in an insurance market that allowed for partial coverage of loss.

In what follows we often refer to “monetary gains and losses” as “monetary returns”. All proofs of major results are relegated to an appendix of this paper. We hope that with this paper, we are able to provide an incremental impetus for further development for decision analysis with linear utility functions for money.

**2. Motivation- Insuring Against Risky Loss:**

Consider a situation with 2 states of nature 1,2, where an agent with initial wealth w > 0 may face a loss of L∈(0, w) units of money in the second SON. Let p∈(0,1) be the probability of loss. Suppose that his utility function for monetary returns in SON i, is a function of the above form with ($u_i^-$, $u_i^+$) being the slopes for losses and gains respectively in SON ‘i’.

There are two ways in which insurance can be introduced in this setting. First is a variant of the traditional textbook setting where we assume $u_2^- > u_1^-$. Even an individual who is not affected by the loss, would react to the news of the loss- by leaning closer towards caution and hence a higher marginal utility of money- than in the absence of such news, however small the difference in the marginal utilities may be. If one hears about frequent bicycle thefts in the neighbourhood that one lives in, then the same person is likely to be concerned more about the safety of his/her bicycle than he/she would be in the absence of such news, regardless of whether the person has been a victim of such theft or not. For an agent with a stake in the loss, the difference gets more pronounced.

In the absence of an insurance policy the expected utility of the agent is $-pu_2^-L$.

An insurance policy that provides complete coverage is available for a premium π which if “actuarily fair” would satisfy π = pL.

The expected utility from buying this policy is $-[(1\text{-}p)\, u_1^- + pu_2^-]\pi = -p[(1\text{-}p)\, u_1^- + pu_2^-]L$.

Since $u_2^- > u_1^- > 0$ and p∈(0,1), $(1\text{-}p)\, u_1^- + pu_2^- < u_2^-$ and so $-p[(1\text{-}p)\, u_1^- + pu_2^-]L > -pu_2^-L$.

Actually, it would be more realistic to consider three SONs: 1-where there is no loss, 2-where there is a loss and the agent “has not” bought the insurance policy and 3- where there is a loss and the agent “has” bought the insurance policy, with $u_2^- > u_3^- > u_1^- > 0$, since having bought the insurance policy, the agent is somewhat more relaxed than what it would be had it not purchased the insurance policy, but since recovering the insurance payment involves some transaction cost (e.g. paper work, etc.) the agent’s disutility from expenditure incurred on buying the premium could be expected to be higher than what it would be had there been no loss.

A second way in which insurance can be introduced in this context, which may be more realistic is to assume that the seller of the insurance policy has recourse to an investment opportunity, which for some r > 0, returns 1 + r units of money for every unit of money invested in the current period. In this case, we can weaken the restriction on the slopes of the utility functions and assume $u_2^- \geq u_1^-$, i.e., allow for $u_2^- = u_1^-$.

In this case an insurance policy that provides complete coverage for a premium π, yields an expected return of (1+ r)π - pL to the seller of the insurance policy which is non-negative if π $\geq \frac{pL}{1+r}$. Since r > 0, $\frac{pL}{1+r} <$ pL, so that the seller of the policy can make a profit by selling it for a premium π∈($\frac{pL}{1+r}$, pL).

In this case, the expected utility from buying this policy for a premium of π is -[(1-p) $u_1^-$ + p$u_2^-$]π and -[(1-p) $u_1^-$ + p$u_2^-$]π > – p$u_2^-$L, since 0 < (1-p) $u_1^-$ + p$u_2^-$ ≤ $u_2^-$ and π < pL.

Now let us consider an agent whose initial monetary wealth is w > 0 and an investible amount I∈(0,w) can either be diversified equally between two-risky investment opportunities or invested entirely in one investment opportunity, with each investment opportunity having a probability p∈(0,1) of failing.

This is a situation where there are three states of nature denoted by 1,2,3 with ($u_i^-$, $u_i^+$) being the slopes for losses and gains respectively in SON 'i' being strictly positive. SON 1 is the situation where neither investment opportunity fails, SON 2 is the situation where 50% of the invested amount is lost and SON 3 is the situation where the entire invested amount is lost.

Suppose 0 < $u_1^-$ < $u_2^-$ < $u_3^-$.

Even if the agent was not an investor, the news of an investment opportunity crashing would very likely have the effect of increasing its disutility of expenditure and such disutility would further increase if it were to hear the news of two investment opportunities crashing simultaneously. In the case of an investor, the effect of such news could only be expected to be more pronounced.

If the agent invests the entire amount I in exactly one investment opportunity, then his expected utility is – p$u_3^-$I.

If the agent spreads his investment opportunity equally between the 2 investment opportunities, then his expected utility is -2p(1-p) $u_2^-\frac{I}{2}$ - $p^2u_3^-$I = -p[(1-p) $u_2^-$ + p$u_3^-$]I.

Since $u_3^-$ > (1-p) $u_2^-$ + p$u_3^-$, we have -p[(1-p) $u_2^-$ + p$u_3^-$]I > – p$u_3^-$I, and hence there is always an incentive for "spreading risks".

**3. The Framework of Analysis:**

Let us now set up the general framework of analysis with state-dependent linear utility functions for monetary returns. For a more general framework of analysis, one may refer to the book by Bonanno (Bonanno 2019).

For some positive integer L ≥ 2, let {1, 2, …, L} denote the finite set of states of nature. As mentioned in the introduction, we will refer to a **state of nature** as **SON** and its plural as **SONs**.

A (column) vector x∈$\mathbb{R}^L$ where for each j∈{1, …, L}, the $j^{th}$ coordinate of x denotes the monetary return in SON j, is said to be a **return vector**.

A (column) vector p∈$\mathbb{R}^L_{++}$ satisfying $\sum_{j=1}^{L} p_j$ = 1, such that for j∈{1, …, L}, $p_j$ is the probability of occurrence of SON j, is a **probability vector.**

Given x, y∈$\mathbb{R}^L$, let $y^T$x denote $\sum_{j=1}^{L} y_j x_j$.

A **portfolio of risky assets** (briefly referred to as **PORA)** is a pair (x, p) where x is a return vector and p is a probability vector. In what follows we will refer to **portfolios of risky assets** as **PORAs**.

Given a PORA (x, p) with X denoting the random monetary return for (x, p) and $\alpha \in \mathbb{R}$, let $\{X = \alpha\}$ denote the event that the realized SON yields a monetary return of $\alpha$, $\{X \leq \alpha\}$ denote the event that the realized SON yields a monetary return less than or equal to $\alpha$, $\{X \geq \alpha\}$ denote the event that the realized SON yields a monetary return greater than or equal to $\alpha$, $\{X < \alpha\}$ denote the event that the realized SON yields a monetary return less than $\alpha$, $\{X > \alpha\}$ denote the event that the realized SON yields a monetary return greater than $\alpha$.

Thus, for all $\alpha \in \mathbb{R}$, Probability of $\{X \leq \alpha\}$ = 1- Probability of $\{X > \alpha\}$

The **expected value** of a PORA (x, p) denoted E(x, p) is $p^T x = \sum_{j=1}^{L} p_j x_j$.

A generalization of the concept of portfolio of risky assets that can be inferred from the solution proposed by Gilboa and Schmeidler (see Gilboa and Schmeidler 1989) as a response to the Ellsberg Paradox is the following.

A **generalized portfolio of risky assets (G-PORA)** is a pair $(\xi, p)$ where $p \in \mathbb{R}^L_{++}$ is a probability vector and for each $j \in \{1, \ldots, L\}$, $\xi_j$ is a non-empty closed and bounded set in $\mathbb{R}$ denoting the set of possible returns in SON j, exactly one from which is realized if SON j occurs.

Unless $\xi_j$ is a singleton, there is no known prior probability distribution over $\xi_j$.

In order to incorporate "ambiguity aversion" one may associate with $(\xi, p)$, the **min portfolio of risky assets (MIN-PORA)** $(\xi^{min}, p)$ which is defined as follows: for each $j \in \{1, \ldots, L\}$, $\xi_j^{min} = \min\{\alpha | \alpha \in \xi_j\}$. For any $x \in \mathbb{R}^L$ satisfying $x_j \in \xi_j$ for all $j \in \{1, \ldots, L\}$, the expected value of the PORA (x, p) can be defined as before, i.e., $E(x, p) = p^T x = \sum_{j=1}^{L} p_j x_j$.

Thus, $E(\xi^{min}, p) = \sum_{j=1}^{L} p_j \xi_j^{min}$.

A **linear utility profile** is an L-tuple $u = (u_1, \ldots, u_L) \in (\mathbb{R}^2_{++})^L$ where for each $j \in \{1, \ldots, L\}$, $u_j$ is a real valued function on $\mathbb{R}$ determined by an ordered pair $(u_j^-, u_j^+) \in \mathbb{R}^2_{++}$ satisfies $u_j^- \geq u_j^+ > 0$, with the interpretation that for all $j \in \{1, \ldots, L\}$ and $\alpha \in \mathbb{R}$, $u_j(\alpha) = u_j^- \min\{\alpha, 0\} + u_j^+ \max\{\alpha, 0\}$ is the (Bernoulli) utility for $\alpha$ units of monetary returns (gains or losses) in SON j.

We allow for the set $\{j | u_j^- = u_j^+\}$ to be $\{1, \ldots, L\}$ or a proper subset of it, including the null set $\phi$.

A similar definition of a linear utility profile has been used in Lahiri (2024) for the purpose of extending the "Arbitrage Theorem" from its usual framework to a one in which state-dependent linear utility functions may allow loss aversion.

Given a linear utility profile $u \in (\mathbb{R}^2_{++})^L$, we will use $u^-$ to denote the vector $(u_1^-, \ldots, u_L^-) \in \mathbb{R}^L_{++}$ and $u^+$ to denote the vector $(u_1^+, \ldots, u_L^+) \in \mathbb{R}^L_{++}$

Given a linear utility profile u and a PORA (x, p) the **expected utility** of (x, p) for u, denoted by Eu(x, p) is $\sum_{j=1}^{L} p_j [u_j^- \min\{x_j, 0\} + u_j^+ \max\{x_j, 0\}]$.

Clearly $Eu(x, p) = p_1(u_1(x_1) - u_2(x_2)) + (p_1 + p_2)(u_2(x_2) - u_3(x_3)) + (p_1 + p_2 + p_3)(u_3(x_3) - u_4(x_4)) + \ldots + (p_1 + \ldots + p_{L-1})(u_{L-1}(x_{L-1}) - u_L(x_L)) + (p_1 + p_2 + \ldots + p_L)u_L(x_L) = \sum_{j=1}^{L-1}(\sum_{k=1}^{j} p_k)\,(u_j(x_j) - u_{j+1}(x_{j+1})) + (\sum_{k=1}^{L} p_k)u_L(x_L) = \sum_{j=1}^{L-1}(\sum_{k=1}^{j} p_k)\,(u_j(x_j) - u_{j+1}(x_{j+1})) + u_L(x_L)$.

Given a linear utility profile u and a PORA (x,p) the **certainty equivalent** of (x,p) for u, denoted by CE(u, x, p) is the scalar that satisfies $\sum_{j=1}^{L} p_j[u_j^- \min\{CE(u,x,p), 0\} + u_j^+ \max\{CE(u,x,p), 0\}] = Eu(x, p)$, i.e. $CE(u,x,p) = \frac{Eu(x,p)}{\sum_{j=1}^{L} p_j u_j^+} = \frac{Eu(x,p)}{p^T u^+}$ if $Eu(x, p) \geq 0$ and $CE(u,x,p) = \frac{Eu(x,p)}{\sum_{j=1}^{L} p_j u_j^-} = = \frac{Eu(x,p)}{p^T u^-}$ if $Eu(x, p) < 0$.

Suppose that (x, p) is a PORA satisfying $x_j < x_{j+1}$ for all $j \in \{1, \ldots, L-1\}$.Then, for all $k \in \{1, \ldots, L-1\}$ and $\alpha, \beta \in (x_k, x_{k+1})$, Probability of $\{X > \alpha\}$ = Probability of $\{X > x_k\}$ = Probability of $\{X > \beta\}$ and Probability of $\{X \leq \alpha\}$ = Probability of $\{X \leq x_k\}$ = Probability of $\{X \leq \beta\}$.

**4. First Order Stochastic Dominance:**

Given two PORAs (x, p) and (y, q) with X denoting the random monetary return for (x, p) and Y denoting the random monetary return for (y, q), we say that (x, p) **stochastically dominates** (y, q) **in the first order**, denoted by $(x, p) >_{FSD} (y,q)$ if for all $\alpha \in \mathbb{R}$, Probability of $\{X > \alpha\} \geq$ Probability of $\{Y > \alpha\}$ and for some $\alpha \in \mathbb{R}$, Probability of $\{X > \alpha\}$ > Probability of $\{Y > \alpha\}$.

The intuitive interpretation of $(x, p) >_{FSD} (y,q)$ is that given any monetary return $\alpha$, the probability that the monetary return from (x, p) is greater than $\alpha$ is greater than or equal to the probability that the monetary return from (y, q) is at greater $\alpha$, and for some monetary return the first probability is strictly greater than the second probability i.e., (x, p) is consistently "more likely" to yield better rewards better than (y, q).

We know that for a linear utility profile and a PORA (x, p), $Eu(x, p) = \sum_{j=1}^{L-1}(\sum_{k=1}^{j} p_k)\,(u_j(x_j) - u_{j+1}(x_{j+1})) + u_L(x_L)$.

**Proposition 1:** Let (x, p) and (x, q) be two PORAs satisfying $x_j < x_{j+1}$ for all $j \in \{1, \ldots, L-1\}$. Then $(x, p) >_{FSD} (x, q)$ if and only if [$Eu(x, p) > Eu(x, q)$ for all linear utility profile u satisfying $u_j(x_j) < u_{j+1}(x_{j+1})$ for all $j \in \{1, \ldots, L-1\}$].

**5. Mean-preserving Spread and Increasing-Concave Linear Utility Profiles:**

**For this section assume** $L \geq 3$.

Given a return vector x satisfying $x_j < x_{j+1}$ for all $j \in \{1, \ldots, L-1\}$, a linear utility profile u is said to be **increasing-concave with respect to** x, if for all $j \in \{1, \ldots, L-1\}$, $u_j(x_j) < u_{j+1}(x_{j+1})$ and for all $i, j, k \in \{1, 2, \ldots, L\}$ with $i < j < k$, $u_j(x_j) > (1-\delta)u_i(x_i) + \delta u_k(x_k)$ where $\delta \in (0,1)$ satisfies $x_j = (1-\delta)x_i + \delta x_k$.

Clearly, $\delta = \frac{x_j - x_i}{x_k - x_i}$ and $0 < x_j - x_i < x_k - x_i$.

Given a return vector x satisfying $x_j < x_{j+1}$ for all $j \in \{1, \ldots, L-1\}$, PORA (x, q) is said to be **obtained by a mean-preserving spread from** PORA (x, p), denoted $(x, p) \rightarrow_{MSP} (x, q)$, if $E(x, p) = E(x, q)$ and there exists $i, j, k \in \{1, 2, \ldots, L\}$ satisfying $i < j < k$ such that $q_i > p_i$, $q_j < p_j$, $q_k > p_k$ and $p_h = q_h$ for $h \in \{1, 2, \ldots, L\} \setminus \{i,j,k\}$.

[$E(x, p) = E(x, q)$ and there exists $i, j, k \in \{1, 2, \ldots, L\}$ satisfying $i < j < k$ such that $q_i > p_i$, $q_j < p_j$, $q_k > p_k$ and $p_h = q_h$ for $h \in \{1, 2, \ldots, L\} \setminus \{i,j,k\}$] <u>if and only if</u> [there exists $i, j, k \in \{1, 2, \ldots, L\}$ satisfying $i < j < k$ such that $q_i > p_i$, $q_j < p_j$, $q_k > p_k$, $p_h = q_h$ for $h \in \{1, 2, \ldots, L\} \setminus \{i,j,k\}$ and $(p_j - q_j)x_j = (q_i - p_i)x_i + (q_k - p_k)x_k$]

[there exists $i, j, k \in \{1, 2, \ldots, L\}$ satisfying $i < j < k$ such that $q_i > p_i$, $q_j < p_j$, $q_k > p_k$ and $p_h = q_h$ for $h \in \{1, 2, \ldots, L\} \setminus \{i,j,k\}$ and $(p_j - q_j)x_j = (q_i - p_i)x_i + (q_k - p_k)x_k$] <u>is equivalent to</u> [there exists $i, j, k \in \{1, 2, \ldots, L\}$ satisfying $i < j < k$ such that $q_i > p_i$, $q_j < p_j$, $q_k > p_k$, $p_h = q_h$ for $h \in \{1, 2, \ldots, L\} \setminus \{i,j,k\}$ and $x_j = \frac{q_i - p_i}{p_j - q_j} x_i + \frac{q_k - p_k}{p_j - q_j} x_k$].

Thus, $(x, p) \rightarrow_{MSP} (x, q)$ <u>if and only if</u> [there exists $i, j, k \in \{1, 2, \ldots, L\}$ satisfying $i < j < k$ such that $q_i > p_i$, $q_j < p_j$, $q_k > p_k$, $p_h = q_h$ for $h \in \{1, 2, \ldots, L\} \setminus \{i,j,k\}$ and $x_j = \frac{q_i - p_i}{p_j - q_j} x_i + \frac{q_k - p_k}{p_j - q_j} x_k$].

**Proposition 2:** Let $(x, p)$ and $(x, q)$ be two PORAs satisfying $x_j < x_{j+1}$ for all $j \in \{1, \ldots, L-1\}$.

(a) <u>If</u> $(x, p) \rightarrow_{MSP} (x, q)$ <u>then</u> [$Eu(x, p) > Eu(x, q)$ for all linear utility profile u which is increasing-concave with respect to x].

(b) <u>If</u> $L = 3$, $p_2 \neq q_2$ and [$Eu(x, p) > Eu(x, q)$ for all linear utility profile u which is increasing-concave with respect to x] <u>then</u> [$(x, p) \rightarrow_{MSP} (x, q)$].

**6. Risk Aversion:**

Given a PORA $(x, p)$, an agent with linear utility profile u is said to be:

(i) **Risk Averse relative to** $(x,p)$ if $E(x, p) > CE(u, x, p)$;

(ii) **Risk Neutral relative to** $(x,p)$ if $E(x, p) = CE(u, x, p)$;

(iii) **Risk Loving/Seeking relative to** $(x, p)$ if $E(x, p) < CE(u, x, p)$.

**Example 1:** Let $L = 2$, $u_1 = (1, 1)$ and let $u_2 = (2, 2)$.

Let $(x, p) = ((2,0), (\frac{1}{2}, \frac{1}{2}))$. Thus, $E(x, p) = 1$.

In this case, $Eu(x,p) = 1$ and $p^T u^+ = \frac{3}{2}$, so that $CE(u, x, p) = \frac{2}{3} < 1 = E(x,p)$.

Thus, the agent is <u>risk averse</u> relative to $((2,0), (\frac{1}{2}, \frac{1}{2}))$.

Now let $(x, p) = ((0,2), (\frac{1}{2}, \frac{1}{2}))$. Once again, $E(x, p) = 1$.

Now, $Eu(x,p) = 2$ and since $p^T u^+ = \frac{3}{2}$, we have $CE(u, x, p) = \frac{4}{3} > 1 = E(x, p)$.

Thus, the same agent is <u>risk loving/seeking</u> relative to $((0,2), (\frac{1}{2}, \frac{1}{2}))$.

Now suppose $(x, p) = ((1,1), (\frac{1}{2}, \frac{1}{2}))$. Once again, $E(x, p) = 1$.

Now, Eu(x,p) = $\frac{3}{2}$ and since $p^T u^+ = \frac{3}{2}$, we have CE(u, x, p) = 1 = E(x, p).

Thus, the same agent is now risk neutral relative to ((1,1), $(\frac{1}{2}, \frac{1}{2})$).

Given a PORA (x,p) and a linear utility profile u, the **risk premium relative to** (x,p) denoted R(u, x, p) = E(x, p) – CE(u, x, p).

Thus, $\sum_{j=1}^{L} p_j u_j\ (E(x,p) - R(u,x,p)) = \sum_{j=1}^{L} p_j u_j\ (CE(u,x,p))$ = Eu(x, p).

If the agent is:

(i) Risk Averse relative to (x, p), then R(u, x, p) > 0;

(ii) Risk Loving/Seeking relative to (x, p), then R(u, x, p) < 0;

(iii) Risk Neutral relative to (x, p), then R(u, x, p) = 0.

Given two linear utility profiles u, v and two PORAs (x, p), (y, q) we say that u relative to (x, p) **is more risk averse than** v relative to (y, q) if R(u, x, p) > R(v, y, q).

**7. Insurance contracts with the possibility of partial coverage:**

As before consider a situation with 2 states of nature 1,2, where an agent with initial wealth w > 0 may face a loss of L∈(0, w) units of money in the second SON. Let p∈(0,1) be the probability of loss. Suppose that the agent's linear utility profile is ($u_1$, $u_2$) is such that with $0 < u_1^- < u_2^-$.

The expected value of the "risk" is -pL

In the absence of an insurance policy the expected utility of the agent is – $pu_2^-$L.

If $CE_1$ is the certainty equivalent in the absence of any insurance policy, then $[(1\text{-p})\, u_1^- + pu_2^-]CE_1 = -pu_2^-L$.

Thus, $CE_1 = \frac{-pu_2^- L}{(1-p)\, u_1^- + pu_2^-} = -pL\frac{u_2^-}{(1-p)\, u_1^- + pu_2^-}$

An insurance policy with a **deductible** d∈[0,L) (i.e., in case of loss, the insurer pays L-d to the agent) is available for a premium π.

Hence the expected profit of the insurer is π - p(L-d).

For the insurer to voluntarily sell the insurance, it must be "**profitable**", i.e., π - p(L-d) ≥ 0.

Thus, profitability is equivalent to the condition - pL ≥ - (π + pd).

The expected value of this policy to the agent is – (π + pd).

The expected utility of the agent from buying this policy is $-(1\text{-p})\, u_1^- \pi - pu_2^-\ (\pi + d) = -[(1\text{-p})\, u_1^- + p\, u_2^-]\pi - pu_2^- d$.

For the agent to voluntarily buy the insurance, it must be the case that $-[(1\text{-p})\, u_1^- + p\, u_2^-]\pi - pu_2^- d \geq -\, pu_2^- L$, i.e., $-\pi - \frac{u_2^-}{(1-p)\, u_1^- + pu_2^-}\, pd \geq CE_1$.

$-\pi - \frac{u_2^-}{(1-p)\, u_1^- + pu_2^-}\, pd = -(\pi + pd) + pd[1-\frac{u_2^-}{(1-p)\, u_1^- + pu_2^-}]$.

Thus the agent will voluntarily buy the insurance policy <u>if and only if</u> $-(\pi + pd) + pd[1-\frac{u_2^-}{(1-p)\, u_1^- + pu_2^-}] \geq CE_1$.

A profit maximizing insurer will choose an **insurance contract**, i.e., a pair ($\pi$, d) that maximizes $\pi - p(L-d)$, subject to $\pi - p(L-d) \geq 0$, $-[(1-p)\, u_1^- + p\, u_2^-]\pi - pu_2^- d \geq -pu_2^- L$ and $d \in [0,L)$.

The above problem is equivalent to choosing a pair ($\pi$, d) that maximizes $\pi + pd$, subject to $\pi + pd \geq pL$, $[(1-p)\, u_1^- + p\, u_2^-]\pi + pu_2^- d \leq pu_2^- L$ and $d \in [0,L)$.

It is easy to see that at an optimal solution, $[(1-p)\, u_1^- + p\, u_2^-]\pi + pu_2^- d = pu_2^- L$.

Thus, $\pi = \frac{pu_2^-(L-d)}{(1-p)\, u_1^- + p\, u_2^-}$.

Thus, $\pi + pd = p[\frac{u_2^-(L-d)}{(1-p)\, u_1^- + p\, u_2^-} + d] = pd[1 - \frac{u_2^-}{(1-p)\, u_1^- + p\, u_2^-}] + \frac{pu_2^- L}{(1-p)\, u_1^- + p\, u_2^-}$.

Since $u_2^- > u_2^1$, we have $\frac{u_2^-}{(1-p)\, u_1^- + p\, u_2^-} > 1$ and hence $1 - \frac{u_2^-}{(1-p)\, u_1^- + p\, u_2^-} < 0$.

Thus, $pd[1 - \frac{u_2^-}{(1-p)\, u_1^- + p\, u_2^-}] + \frac{pu_2^- L}{(1-p)\, u_1^- + p\, u_2^-}$ is maximized at d = 0, thereby implying $\pi = \frac{pu_2^- L}{(1-p)\, u_1^- + p\, u_2^-}$.

Since $\frac{pu_2^- L}{(1-p)\, u_1^- + p\, u_2^-} = (\frac{u_2^-}{(1-p)\, u_1^- + p\, u_2^-})pL$ and $\frac{u_2^-}{(1-p)\, u_1^- + p\, u_2^-} > 1$, we have $\pi > pL$. Since d = 0, $\pi + pd > pL$.

Hence, the optimal contract is the pair $(\frac{pu_2^- L}{(1-p)\, u_1^- + p\, u_2^-}, 0)$, with the "expected profit of the insurer" being $\frac{pu_2^- L}{(1-p)\, u_1^- + p\, u_2^-} - pL = pL(\frac{u_2^- - (1-p)u_1^- - p\, u_2^-}{(1-p)\, u_1^- + p\, u_2^-}) = \frac{p(1-p)(u_2^- - u_1^-)L}{(1-p)u_1 + pu_2^-} > 0$.

**Note:** $\pi = \frac{pu_2^-(L-d)}{(1-p)\, u_1^- + p\, u_2^-}$ implies $-\pi - \frac{u_2^-}{(1-p)\, u_1^- + p\, u_2^-}\, pd = -\frac{pu_2^- L}{(1-p)\, u_1^- + p\, u_2^-} = CE_1$.

We know that $-\pi - \frac{u_2^-}{(1-p)\, u_1^- + p\, u_2^-}\, pd = -(\pi + pd) + pd[1-\frac{u_2^-}{(1-p)\, u_1^- + p\, u_2^-}]$.

Thus, at an optimal solution $-(\pi + pd) + pd[1-\frac{u_2^-}{(1-p)\, u_1^- + p\, u_2^-}] = CE_1$.

"**Strict Profitability**" is equivalent to the condition $-pL > -(\pi + pd)$ which now reduces to

$-pL + pd[1-\frac{u_2^-}{(1-p)\, u_1^- + p\, u_2^-}] > CE_1 = -\frac{pu_2^- L}{(1-p)\, u_1^- + p\, u_2^-}$.

Thus **strict profitability** is equivalent to $-pL[1-\frac{u_2^-}{(1-p)\, u_1^- + p\, u_2^-}] + pd[1-\frac{u_2^-}{(1-p)\, u_1^- + p\, u_2^-}] > 0$, i.e. $p(d-L)[1-\frac{u_2^-}{(1-p)\, u_1^- + p\, u_2^-}] \geq 0$.

Since $d \in [0, L)$, this is possible <u>if and only if</u> $1-\frac{u_2^-}{(1-p)\, u_1^- + p\, u_2^-} < 0$, i.e. $1 < \frac{u_2^-}{(1-p)\, u_1^- + p\, u_2^-}$

Multiplying throughout by pL which is strictly positive, we get $1 < \frac{u_2^-}{(1-p)\, u_1^- + p\, u_2^-}$ if and only if $pL < \frac{u_2^-}{(1-p)\, u_1^- + p\, u_2^-} pL$, the latter being equivalent to $- \frac{u_2^-}{(1-p)\, u_1^- + p\, u_2^-} pL < -pL$.

Since $CE_1 = - \frac{u_2^-}{(1-p)\, u_1^- + p\, u_2^-}$ pL and -pL is the expected value of the "risk", the agent is risk averse relative to ((-L, 0), (p, 1-p)).

Thus "**Strict Profitability**" is equivalent to the requirement that the agent is risk averse relative to ((-L, 0), (p, 1-p)).

Let us now consider the somewhat more realistic situation with three SONs: 1-where there is no loss, 2- where there is a loss and the agent "has not" bought the insurance policy and 3- where there is a loss and the agent "has" bought the insurance policy, with $u_2^- > u_3^- > u_1^- > 0$.

Then, the expected utility of the agent from buying this policy is $-(1-p)\, u_1^- \pi - p u_3^- (\pi + d) = -[(1-p)\, u_1^- + p u_3^-]\pi - p u_3^- d$.

Since $u_2^- > u_3^-$, $- (1-p) u_1^- \pi - p u_3^- (\pi + d) > - (1-p)\, u_1^- \pi - p u_2^- (\pi + d)$.

A profit maximizing insurer will choose an **insurance contract**, i.e., a pair (π, d) that maximizes π - p(L-d) subject to π - p(L-d) ≥ 0, $-[(1-p)\, u_1^- + p u_3^-]\pi - p u_3^- d \geq - p u_2^- L$ and d∈[0,L].

The above problem is equivalent to choosing a pair (π, d) that maximizes π + pd subject to π + pd ≥ pL, $[(1-p)\, u_1^- + p u_3^-]\pi + p u_3^- d \leq p u_2^- L$ and d∈[0,L].

It is easy to see that at an optimal solution, $[(1-p) u_1^- + p u_3^-]\pi + p u_3^- d = p u_2^- L$.

Thus, $\pi = \frac{p(u_2^- L - u_3^- d)}{(1-p)\, u_1^- + p u_3^-}$.

Thus, $\pi + pd = p[\frac{(u_2^- L - u_3^- d)}{(1-p)\, u_1^- + p u_3^-} + d] = pd[1 - \frac{u_3^-}{(1-p) u_1^- + p u_3^-}] + \frac{p u_2^- L}{(1-p) u_1^- + p u_3^-}$.

Since $u_3^- > u_3^1$, we have $\frac{u_3^-}{(1-p) u_1^- + p u_3^-} > 1$ and hence $1 - \frac{u_3^-}{(1-p) u_1^- + p u_3^-} < 0$.

Thus, $pd[1 - \frac{u_3^-}{(1-p) u_1^- + p u_3^-}] + \frac{p u_2^- L}{(1-p) u_1^- + p u_3^-}$ is maximized at d = 0, thereby implying $\pi = \frac{p u_2^- L}{(1-p) u_1^- + p u_3^-} > \frac{p u_3^- L}{(1-p) u_1^- + p u_3^-}$, since $u_2^- > u_3^-$.

Since d = 0, $\pi + pd = \frac{p u_2^- L}{(1-p) u_1^- + p u_3^-} > pL$.

Hence, the optimal contract is the pair $(\frac{p u_2^- L}{(1-p) u_1^- + p u_3^-}, 0)$, with the "expected profit of the insurer" being $\frac{p u_2^- L}{(1-p) u_1^- + p u_3^-} - pL > \frac{p u_3^- L}{(1-p) u_1^- + p u_3^-} - pL$.

Thus, the expected profit of the insurer is higher in this more realistic situation than in the earlier situation.

**Acknowledgment:** I would like to that Itzhak Gilboa for comments and queries about an earlier version. I would also like to apologise to R. Chandrasekar (Sekar) for claiming his valuable respite from "real-world finance" during weekends, to glance through the work of "dreamers(?)". I hope practitioners of finance will find less dreaming here than in the eternal truth here: "https://www.youtube.com/watch?v=hWWvTTbCGTI".

## Appendix

**Proof of Proposition 1:** Eu(x, p) – Eu(x, q) = $[\sum_{j=1}^{L-1}(\sum_{k=1}^{j} p_k)\,(u_j\,(x_j) - u_{j+1}(x_{j+1})\,) + u_L(x_L)] - [\sum_{j=1}^{L-1}(\sum_{k=1}^{j} q_k)\,(u_j\,(x_j) - u_{j+1}(x_{j+1})\,) + (\sum_{k=1}^{L} q_k)\, u_L(x_L)] = \sum_{j=1}^{L-1}(\sum_{k=1}^{j} p_k - \sum_{k=1}^{j} q_k)\,(u_j\,(x_j) - u_{j+1}(x_{j+1})\,)$, since $\sum_{k=1}^{L} p_k = 1 = \sum_{k=1}^{L} q_k$.

Suppose (x, p) $>_{FSD}$ (x, q). Then, $\sum_{k=1}^{j} p_k - \sum_{k=1}^{j} q_k \leq 0$ for all $j \in \{1, \ldots, L\}$, with strict inequality for at least one $j \in \{1, \ldots, L-1\}$, since $\sum_{k=1}^{L} p_k = 1 = \sum_{k=1}^{L} q_k$.

If u is a linear utility profile satisfying $u_j(x_j) < u_{j+1}(x_{j+1})$ for all $j \in \{1, \ldots, L-1\}$, then $\sum_{j=1}^{L-1}(\sum_{k=1}^{j} p_k - \sum_{k=1}^{j} q_k)\,(u_j\,(x_j) - u_{j+1}(x_{j+1})\,) > 0$.

Thus, Eu(x, p) – Eu(x, q) > 0, i.e., Eu(x, p) > Eu(x, q).

Now suppose that it is not the case that (x, p) $>_{FSD}$ (x, q).

Thus, $\{j \in \{1, \ldots, L-1\} |\, \sum_{k=1}^{j} p_k - \sum_{k=1}^{j} q_k > 0\} \neq \phi$. Let $\eta = \min\{\sum_{k=1}^{j} p_k - \sum_{k=1}^{j} q_k |\, \sum_{k=1}^{j} p_k - \sum_{k=1}^{j} q_k > 0\}$.

Let $u_1 = (1, 1)$. Having defined $u_j = (u_j^-, u_j^+) \in \mathbb{R}_{++}^2$ satisfying $u_j^- = u_j^+ > 0$, let $u_{j+1} = (u_{j+1}^-, u_{j+1}^+) \in \mathbb{R}_{++}^2$ satisfying $u_{j+1}^- = u_{j+1}^+ > 0$ be such that $u_{j+1}(x_{j+1}) - u_j(x_j) = \frac{2}{\eta}$ if $\sum_{k=1}^{j} p_k - \sum_{k=1}^{j} q_k > 0$ and $\frac{1}{2L} > u_{j+1}(x_{j+1}) - u_j(x_j) > 0$, otherwise. This is possible, since $x_{j+1} > x_j$ implies that it is not possible for both $x_{j+1}$ and $x_j$ to be zero.

Thus, $Eu(x, p) - Eu(x, q) = -\frac{2}{\eta}\sum_{h\in\{\{j\in\{1,\ldots,L-1\}|\ \sum_{k=1}^{j}p_k-\sum_{k=1}^{j}q_k>0\}}\sum_{k=1}^{h}p_k-\sum_{k=1}^{h}q_k+$
$\sum_{h\in\left\{\{j\in\{1,\ldots,L-1\}|\sum_{k=1}^{j}p_k-\sum_{k=1}^{j}q_k\le 0\right\}}\left(\sum_{k=1}^{h}p_k-\sum_{k=1}^{h}q_k\right)\left(u_h(x_h)-u_{h+1}(x_{h+1})\right)=$
$-\frac{2}{\eta}\sum_{h\in\{\{j\in\{1,\ldots,L-1\}|\ \sum_{k=1}^{j}p_k-\sum_{k=1}^{j}q_k>0\}}\sum_{k=1}^{h}p_k-\sum_{k=1}^{h}q_k+$
$\sum_{h\in\left\{\{j\in\{1,\ldots,L-1\}|\sum_{k=1}^{j}p_k-\sum_{k=1}^{j}q_k\le 0\right\}}\left(\sum_{k=1}^{h}q_k-\sum_{k=1}^{h}p_k\right)\left(u_{h+1}(x_{h+1})-u_h(x_h)\right)\le -2+(L-1)\frac{1}{2L}$
$\le -2+\frac{1}{2}=-\frac{3}{2}<0$.

Thus, [$Eu(x, p) > Eu(x, q)$ for all linear utility profile u satisfying $u_j(x_j) < u_{j+1}(x_{j+1})$ for all $j\in\{1, \ldots, L-1\}$] implies $(x, p) >_{FSD} (x, q)$. Q.E.D.

**Proof of Proposition 2:** (a) Suppose $(x, p) \rightarrow_{MSP} (x, q)$ and let u be an increasing-concave linear utility profile with respect to x.

Hence, there exists i, j, $k\in\{1, 2, \ldots, L\}$ satisfying $i < j < k$ such that $q_i > p_i$, $q_j < p_j$, $q_k > p_k$ and $p_h = q_h$ for $h\in\{1, 2, \ldots, L\}\backslash\{i,j,k\}$and $(p_j-q_j)x_j = (q_i- p_i)x_i + (q_k - p_k)x_k$] is equivalent to [there exists i, j, $k\in\{1, 2, \ldots, L\}$ satisfying $i < j < k$ such that $q_i > p_i$, $q_j < p_j$, $q_k > p_k$, $p_h = q_h$ for $h\in\{1, 2, \ldots, L\}\backslash\{i,j,k\}$ and $x_j = \frac{q_i-p_i}{p_j-q_j}x_i + \frac{q_k-p_k}{p_j-q_j}x_k$.

However, $x_j = (1-\delta)x_i + \delta x_k$ where $\delta = \frac{x_j-x_i}{x_k-x_i}\in(0, 1)$.

Further $p_i + p_j + p_k = q_i + q_j + q_k$ implies $p_j - q_j = (q_i - p_i) + (q_k - p_k)$.

Thus, $\frac{q_i-p_i}{p_j-q_j} + \frac{q_k-p_k}{p_j-q_j} = 1$ with $\frac{q_i-p_i}{p_j-q_j} > 0$ and $\frac{q_k-p_k}{p_j-q_j} > 0$.

Hence, $\frac{q_k-p_k}{p_j-q_j} = \delta$ and $\frac{q_i-p_i}{p_j-q_j} = 1-\delta$.

Since u is increasing-concave $u_j(x_j) > (1-\delta)u_i(x_i) + \delta u_k(x_k)$.

Thus, $(p_j - q_j)u_j(x_j) > (q_i - p_i)u_i(x_i) + (q_k - p_k)u_k(x_k)$, i.e., $p_iu_i(x_i) + p_ju_j(x_j) + p_ku_k(x_k) > q_iu_i(x_i) + q_ju_j(x_j) + q_ku_k(x_k)$.

Since, $p_h = q_h$ for $h\in\{1, 2, \ldots, L\}\backslash\{i,j,k\}$, we get $Eu(x, p) > Eu(x, q)$.

(b) Now suppose $L = 3$ and $x_1 < x_2 < x_3$ and $p_2 \neq q_2$.

We have $p_1 + p_2 + p_3 = q_1 + q_2 + q_3 = 1$.

Suppose, $E(x, p) = E(x, q)$. Thus, $p_1x_1 + p_2x_2 + p_3x_3 = q_1x_1 + q_2x_2 + q_3x_3$.

Suppose, $Eu(x, p) > Eu(x, q)$ for all linear utility profiles satisfying $u_1(x_1) < u_2(x_2) < u_3(x_3)$ and $u_2(x_2) > (1-\delta)u_1(x_1) + \delta u_3(x_3)$, where $x_2 = (1-\delta)x_1 + \delta x_3$.

Since $p_2 - q_2 \neq 0$, $(p_2-q_2)x_2 = (q_1 - p_1)x_1 + (q_3 - p_3)x_3$ implies $x_2 = \frac{q_1-p_1}{p_2-q_2}x_1 + \frac{q_3-p_3}{p_2-q_2}x_3 = \frac{q_1-p_1}{p_2-q_2}x_1 + \frac{(1-q_1-q_2)-(1-p_1-p_2)}{p_2-q_2}x_3 = \frac{q_1-p_1}{p_2-q_2}x_1 + \frac{(p_2-q_2)-(q_1-p_1)}{p_2-q_2}x_3 = x_3 - \frac{q_1-p_1}{p_2-q_2}(x_3 - x_1)$, i.e., $x_2 = x_3 - \frac{q_1-p_1}{p_2-q_2}(x_3 - x_1)$.

$x_2 < x_3$ and $x_3 > x_1$ implies $\frac{q_1-p_1}{p_2-q_2} > 0$.

Similarly, $x_2 = \frac{q_1 - p_1}{p_2 - q_2}x_1 + \frac{q_3 - p_3}{p_2 - q_2}x_3 = \frac{(1 - q_2 - q_3) - (1 - p_2 - p_3)}{p_2 - q_2}x_1 + \frac{q_3 - p_3}{p_2 - q_2}x_3 = \frac{(p_2 - q_2) - (q_3 - p_3)}{p_2 - q_2}x_1 + \frac{q_3 - p_3}{p_2 - q_2}x_3 = x_1 + \frac{q_3 - p_3}{p_2 - q_2}(x_3 - x_1)$.

$x_2 > x_1$ and $x_3 > x_1$ implies $\frac{q_3 - p_3}{p_2 - q_2} > 0$.

Thus, $x_2 = \frac{q_1 - p_1}{p_2 - q_2}x_1 + \frac{q_3 - p_3}{p_2 - q_2}x_3$, $x_2 = (1-\delta)x_1 + \delta x_3$, $\delta > 0$, $1- \delta > 0$, $\frac{q_3 - p_3}{p_2 - q_2} > 0$, $\frac{q_1 - p_1}{p_2 - q_2} > 0$ and $x_1 < x_2 < x_3$ implies $\delta = \frac{q_3 - p_3}{p_2 - q_2}$ and $1- \delta = \frac{q_1 - p_1}{p_2 - q_2}$.

Thus, $u_2(x_2) > \frac{q_1 - p_1}{p_2 - q_2} u_1(x_1) + \frac{q_3 - p_3}{p_2 - q_2} u_3(x_3)$.

If $p_2 < q_2$, then $(p_2 - q_2)u_2(x_2) < (q_1 - p_1)u_1(x_1) + (q_3 - p_3)u_3(x_3)$ and thus, $Eu(x, p) = p_1u_1(x_1) + p_2u_2(x_2) + p_3u_3(x_3) < q_1u_1(x_1) + q_2u_2(x_2) + q_3u_3(x_3) = Eu(x, q)$, leading to a contradiction.

Thus, it must be the case that $p_2 > q_2$.

Hence, $\frac{q_1 - p_1}{p_2 - q_2} > 0$ implies $q_1 > p_1$ and $\frac{q_3 - p_3}{p_2 - q_2} > 0$ implies $q_3 > p_3$.

Thus, we have $(x, p) \rightarrow_{MSP} (x, q)$. Q.E.D.

**Note:** The proof of part (b) in Proposition 2, can very likely be extended to L > 3.